\documentclass[twocolumn,10pt]{IEEEtran}
\usepackage{amsmath, graphics,amssymb,epsfig,subfigure}

\usepackage{cite}
\usepackage{url}
\usepackage{color,soul}

\def\BibTeX{{\rm B\kern-.05em{\sc i\kern-.025em b}\kern-.08em
    T\kern-.1667em\lower.7ex\hbox{E}\kern-.125emX}}

\newcommand{\bea}{\begin{eqnarray}}
\newcommand{\eea}{\end{eqnarray}}
\usepackage{graphicx,amsmath}

\makeatletter
\def\hlinewd#1{%
\noalign{\ifnum0=`}\fi\hrule \@height #1 \futurelet \reserved@a\@xhline}
\newcommand{\hthickline}{\hlinewd{.8pt}}

\makeatother
\begin{document}

\title{\huge{Joint Design of Fronthauling and Hybrid Beamforming for Downlink C-RAN Systems}}

\author{\IEEEauthorblockN{\normalsize{Jaein Kim, \textit{Student Member}, \textit{IEEE}, Seok-Hwan Park, \textit{Member}, \textit{IEEE}, \\
Osvaldo Simeone, \textit{Fellow}, \textit{IEEE}, Inkyu Lee, \textit{Fellow}, \textit{IEEE}, and\\
Shlomo Shamai (Shitz), \textit{Life Fellow}, \textit{IEEE}}\\
 }
\thanks{
This work was supported in part by the National Research Foundation (NRF) through the Ministry of Science, ICT and Future Planning, Korean Government, under Grant 2017R1A2B3012316.
The work of S.-H. Park was supported by the NRF grant funded by the Korea government [NRF-2018R1D1A1B07040322].
The work of O. Simeone was supported by the European Research Council (ERC) under the European Union Horizon 2020 research and innovative programme (grant agreement No 725731).
The work of S. Shamai has been supported by the European Union's Horizon 2020, grant agreement No. 694630.
This paper was presented in part at the IEEE SPAWC 2017, Sapporo, Japan, July 2017 \cite{JKim}.

J. Kim and I. Lee are with the School of Electrical Engineering, Korea University, Seoul 02841, Korea (e-mail: \{kji\_07, inkyu\}@korea.ac.kr).

S.-H. Park is with the Division of Electronic Engineering, Chonbuk National University, Jeonju 54896, Korea (e-mail: seokhwan@jbnu.ac.kr).

O. Simeone is with the Department of Informatics, King's College London, London WC2R 2LS, UK (email: osvaldo.simeone@kcl.ac.uk).

S. Shamai (Shitz) is with the Department of Electrical Engineering, Technion, Haifa, 32000, Israel (e-mail: sshlomo@ee.technion.ac.il).}
}\maketitle \thispagestyle{empty}
\begin{abstract}

Hybrid beamforming is known to be a cost-effective and wide-spread solution for a system with large-scale antenna arrays.
This work studies the optimization of the analog and digital components of the hybrid beamforming solution for remote radio heads (RRHs) in a downlink cloud radio access network (C-RAN) architecture.
Digital processing is carried out at a baseband processing unit (BBU) in the \textquotedblleft cloud\textquotedblright~and the precoded baseband signals are quantized prior to transmission to the RRHs via finite-capacity fronthaul links.
In this system, we consider two different channel state information (CSI) scenarios:
1) ideal CSI at the BBU 2) imperfect effective CSI.
Optimization of digital beamforming and fronthaul quantization strategies at the BBU as well as analog radio frequency (RF) beamforming at the RRHs is a coupled problem, since the effect of the quantization noise at the receiver depends on the precoding matrices.
The resulting joint optimization problem is examined with the goal of maximizing the weighted downlink sum-rate and the network energy efficiency.
Fronthaul capacity and per-RRH power constraints are enforced along with constant modulus constraint on the RF beamforming matrices.
For the case of perfect CSI, a block coordinate descent scheme is proposed based on the weighted minimum-mean-square-error approach by relaxing the constant modulus constraint of the analog beamformer.
Also, we present the impact of imperfect CSI on the weighted sum-rate and network energy efficiency performance, and the algorithm is extended by applying the sample average approximation.
Numerical results confirm the effectiveness of the proposed scheme and show that the proposed algorithm is robust to estimation errors.

\end{abstract}


\section{Introduction} \label{sec:intro}
Massive multiple-input and multiple-output (MIMO) has been regarded as a promising technology for future wireless systems owing to its potential of improving both spectral and energy efficiency (EE) with simple signal processing \cite{LL:14}.
This is enabled by the fact that the channel vectors for different users become orthogonal when the number of transmit antennas grows to infinity.
However, with massive MIMO arrays, it is generally impractical to equip every antenna of a large array with a radio frequency (RF) chain due to hardware limitations \cite{AFM:17}.
Hybrid beamforming techniques, whereby the beamforming process consists of a low-dimensional digital beamforming followed by analog RF beamforming, has emerged as an effective means to address this problem (see, e.g., \cite{OEAy:14,LLiang:14,LLi:14,AAl:15,FSo:16,XYu:16,CSL:16,SPark:17,ALiu:18,TO:18}).
Both analog and digital components are typically designed separately and locally for a base station (BS) \cite{AFM:17}.

In a cloud radio access network (C-RAN) architecture, the baseband signal processing functionalities of multiple BSs are migrated to a baseband processing unit (BBU) in the \textquotedblleft cloud\textquotedblright, while RF functionalities are implemented at distributed remote radio heads (RRHs).
Therefore, in the C-RAN architecture with large antenna arrays at the RRHs, digital precoding across multiple RRHs can be carried out at the BBU, while RF beamforming is performed locally at each RRH.
The design problem becomes more challenging by the capacity limitations of the fronthaul links that connect the BBU to the RRHs.

In the downlink of C-RAN, the BBU performs joint encoding and precoding of the messages intended for user equipments (UEs), and then the produced baseband signals are quantized and compressed prior to being transferred to the RRH via fronthaul links.
The design of precoding and fronthaul compression strategies has been studied in \cite{OS:09,SHP:13,SHPark:14}.
Specifically, the authors in \cite{OS:09} considered the standard point-to-point fronthaul compression strategies.
In \cite{SHP:13,SHPark:14}, the authors investigated multivariate fronthaul compression.

In this work, we study the application of hybrid beamforming to the C-RAN architecture.
We tackle the problem of jointly optimizing digital baseband beamforming and fronthaul compression strategies at the BBU along with RF beamforming at the RRHs with the goal of maximizing the weighted downlink sum-rate and the network EE.
Fronthaul capacity and per-RRH transmit power constraints are imposed, as well as constant modulus constraint on the RF beamforming matrices which consist of analog phase shifters \cite{AFM:17}.

The limited number of RF chains determines the capability of the BBU to acquire channel state information (CSI) through conventional uplink training based on the time division duplex (TDD) operation.
In particular, during the uplink training, the received baseband signal depends on a RF beamforming matrix, and hence instantaneous CSI is unavailable when designing the RF beamforming matrices.
To address this limitation, the RF beamforming matrices are computed based on the second-order statistics of the downlink channel vectors, and the digital beamforming and fronthaul compression strategies are adaptive to the estimated effective channel.

\subsection{Related Work}
A hybrid beamforming design has been investigated in \cite{OEAy:14,LLiang:14,LLi:14,AAl:15,FSo:16,XYu:16,CSL:16,SPark:17,ALiu:18,TO:18}.
Specifically, in \cite{OEAy:14}, a point-to-point hybrid precoding and combining algorithm was proposed that uses orthogonal matching pursuit for millimeter-wave (mmWave) systems.
The authors in \cite{LLi:14} provided a low-complexity hybrid beamforming scheme to achieve sum-rate performance close to that of the zero-forcing (ZF) method for the downlink of multi-user multiple-input single-output (MISO) systems.
In this case, each RF beamforming vector for a user was determined by projecting the downlink channel onto the feasible RF space with low-dimensional ZF digital beamforming.
In addition, for multi-user MIMO mmWave systems, a limited feedback hybrid beamforming scheme was presented in \cite{AAl:15}.
The work in \cite{FSo:16} proved that in hybrid beamforming, the number of RF chains needs to be twice the number of data streams to achieve sum-rate performance equal to that of fully digital beamforming.
Also, the authors in \cite{FSo:16} considered a design of the hybrid beamforming to maximize spectral efficiency for point-to-point MIMO and multi-user MISO scenarios.

Most of works on hybrid beamforming in \cite{OEAy:14,LLiang:14,LLi:14,AAl:15,FSo:16,XYu:16,CSL:16} have assumed full CSI.
However, it is difficult to estimate the channel vectors across all antenna elements, since the estimation operates in the low-dimensional baseband downlink obtained after RF beamforming.
To address this issue, in \cite{SPark:17} and \cite{ALiu:18}, the RF beamforming matrix was determined by using the long-term CSI, while the digital beamformer was designed based on the low-dimensional effective channel.
Recently, a design of hybrid beamforming for C-RAN systems has been studied in \cite{QH} and \cite{Jpark}.
The authors in \cite{QH} provided a two-stage algorithm that only demands low-dimensional effective CSI.
In \cite{Jpark}, a RF beamforming was computed based on a weighted sum of second-order channel statistics and the size of RF and digital beamforming matrices was determined in order to maximize the large-scale approximated sum-rate with regularized ZF digital beamforming.

Furthermore, the EE maximization problem in C-RAN has been studied in \cite{PL:18,TTV:18,KGN:18}.
In \cite{PL:18}, the authors considered a joint design of beamforming, virtual computing resources, RRH selection, and RRH-UE association in a limited fronthaul C-RAN, and a global optimization algorithm and a low-complexity method were presented.
Also, for both single-hop and multi-hop C-RAN scenarios, the problem of EE maximization under both data-sharing and compression-based fronthaul strategies was addressed in \cite{TTV:18}.
The authors in \cite{KGN:18} took into account a realistic power consumption model which is dependent on the data rate and dynamic power amplifier.

\subsection{Main Contributions, Paper Organization and Notation}

The main contributions of this paper are as follows:
\begin{itemize}
\item For downlink C-RAN with hybrid analog-digital antenna arrays, we investigate the joint design of fronthauling and hybrid beamforming with the goal of maximizing the weighted sum-rate (WSR) and the network EE.
\item For the case of perfect CSI, we first decompose the problem into two sub-problems of the RF beamforming and the digital processing, which turn out to be non-convex, and then we propose an iterative algorithm based on weighted minimum-mean-square-error (WMMSE) approach by relaxing constant modulus constraint.
    Also, for the imperfect CSI case, we extend the solution with the perfect CSI by applying the sample average approximation (SAA) \cite{ALiu:16}.
\item Extensive numerical results are provided to validate the effectiveness of the proposed algorithm.
In addition, in the presence of channel estimation errors, we show the robustness of the proposed scheme.
\end{itemize}

The paper is organized as follows:
In Sec. \ref{sec:system}, we present the system model for the downlink of a C-RAN with hybrid digital and analog processing and finite-capacity fronthaul links.
In Sec. \ref{sec:design_perfect}, for the case of perfect CSI, we describe the problems of WSR maximization and network EE maximization, and an iterative algorithm to tackle the problems is proposed.
Sec. \ref{sec:design_imperfect} discusses the problem of CSI estimation in a TDD system.
In addition, we introduce an uplink channel training method and provide a RF beamforming matrix design and the digital strategies.
Numerical results are illustrated in Sec. \ref{sec:results}.
The paper is closed with the conclusion in Sec. \ref{sec:conclusion}.

\begin{table}
\caption{Definition of variables.}
\begin{tabular} {|c|c|c|c|}
  \hline
   Variable & Definition
   \\\hline\hline
   $\mathbf{V}_{R,i}$ & RF beamforming matrix at the $i$th RRH\\\hline
   $\mathbf{v}_{D,k}$ & Digital beamforming vector for the $k$th UE\\\hline
   $\mathbf{\Omega}_i$ & Quantization noise covariance matrix at the $i$th RRH\\\hline
   $u_k$ & MMSE receiver at the $k$th UE\\\hline
   $\tilde{w}_k$ & Weight variable for the $k$th UE\\\hline
   $\mathbf{\Sigma}_i$ & Auxiliary matrix at the $i$th RRH\\\hline
   $\rho$ & Auxiliary variable for power consumption\\\hline
   $\alpha_t$ & Additional variable at the $t$th instantaneous channel sample\\\hline
   $\beta_t$ & Auxiliary variable at the $t$th instantaneous channel sample\\\hline
  \end{tabular}
\end{table}

Throughout this paper, boldface uppercase, boldface lowercase and normal letters indicate matrices, vectors and scalars, respectively.
The operators $(\cdot)^T$, $(\cdot)^H$, $\mathbb{E}(\cdot)$, $\text{det}(\cdot)$ and $\text{tr}(\cdot)$ represent transpose, conjugate transpose, expectation, determinant and trace, respectively.
A circularly symmetric complex Gaussian distribution with mean $\mathbf{\mu}$ and covariance matrix $\mathbf{R}$ is denoted by $\mathcal{CN}(\mathbf{\mu},\mathbf{R})$.
The set of all $M\times N$ complex matrices is defined as $\mathbb{C}^{M\times N}$.
$\mathbf{I}_d$ represents an identity matrix of size $d$.
$\otimes$ stands for the Kronecker product.
The variables used in this paper are summarized in Table I.

\section{System Model} \label{sec:system}

\begin{figure}
\begin{center}
\includegraphics[width=3.3 in]{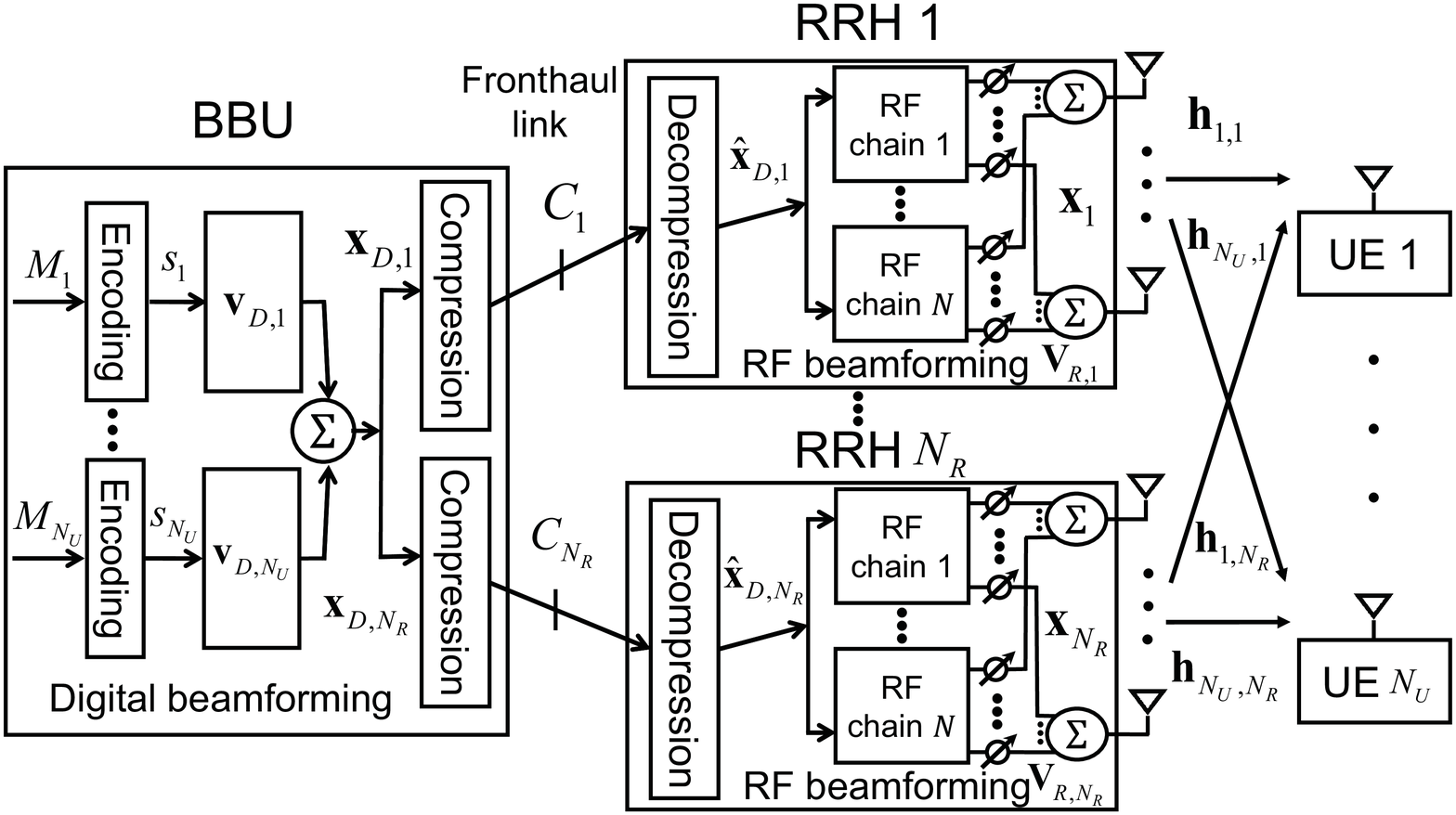}
\end{center}
\vspace{-5mm}
\caption{Illustration of the downlink of a C-RAN with hybrid digital and analog processing.}
\label{figure:Fig1}
\end{figure}

As illustrated in Fig. \ref{figure:Fig1}, we consider the downlink of a C-RAN system in which a BBU communicates with $N_U$ single-antenna UEs through $N_R$ RRHs, each equipped with $M$ transmit antennas.
We assume that the $i$th RRH is connected to the BBU via an error-free digital fronthaul link of capacity $C_i$ bps/Hz \cite{SHP:16,JY:16}, and each RRH is equipped with $N\leq M$ RF chains due to cost limitations.
This implies that fully digital beamforming across $M$ transmit antennas of each RRH is not enabled \cite{FSo:16,LLiang:14,OEAy:14,LLi:14,AAl:15} and thus hybrid analog-digital solutions are in order.
For convenience, we define the sets $\mathcal{R}\triangleq\{1,\cdots,N_R\}$, $\mathcal{K}\triangleq\{1,\cdots,N_U\}$, $\mathcal{M}\triangleq\{1,\cdots,M\}$ and $\mathcal{N}\triangleq\{1,\cdots,N\}$.
We assume that user scheduling is predetermined and hence all the $N_U$ UEs are active.
For rate allocation, the priority among $N_U$ active UEs can be controlled by adjusting the weights of the weighted sum-rate in Sec. \ref{sec:SR_perfect}.
The transmit-side baseband processing is centralized at the BBU based on CSI reported by the RRHs on the fronthaul links.
Assuming a TDD operation, each RRH obtains its local CSI by means of uplink channel training \cite{Guvensen,LZhao}.

\subsection{Channel Model} \label{sec:channel-model}
For the downlink channel from the RRHs to the UEs, we adopt a frequency-flat fading channel model such that the received signal $y_k$ at the $k$th UE is given as
\bea
y_k=\sum_{i\in\mathcal{R}}\mathbf{h}_{k,i}^H\mathbf{x}_i+z_k=\mathbf{h}_k^H\mathbf{x}+z_k,\label{y_k}
\eea
where $\mathbf{x}_i\in\mathbb{C}^{M\times 1}$ is the transmitted signal of the $i$th RRH subject to the transmit power constraint $\mathbb{E}\|\mathbf{x}_i\|^2\leq P_i$, $\mathbf{h}_{k,i}\in\mathbb{C}^{M\times 1}$ equals the channel vector from the $i$th RRH to the $k$th UE, which is distributed as $\mathbf{h}_{k,i}\sim\mathcal{CN}(\mathbf{0},\mathbf{R}_{k,i})$ with $\mathbf{R}_{k,i}$ being the transmit-side correlation, $\mathbf{x}=[\mathbf{x}_1^H\cdots\mathbf{x}_{N_R}^H]^H\in\mathbb{C}^{MN_R\times 1}$ represents the signal vector transmitted by all RRHs, $\mathbf{h}_k=[\mathbf{h}_{k,1}^H\cdots\mathbf{h}_{k,N_R}^H]^H\in\mathbb{C}^{MN_R\times 1}$ indicates the channel vector from all RRHs to the $k$th UE, and $z_k\sim\mathcal{CN}(0,\sigma_D^2)$ denotes the additive noise at the $k$th UE.

\subsection{Digital Beamforming and Fronthaul Compression}
We define the message intended for the $k$th UE as $M_k\in\{1,\cdots,2^{nR_k}\}$, where $n$ stands for the coding block length and $R_k$ is the rate of $M_k$.
The BBU encodes the message $M_k$ into the baseband signal $s_k\sim\mathcal{CN}(0,1)$ for $k\in\mathcal{K}$ using a standard random Gaussian channel code.
Then, in order to manage inter-UE interference, the signals $\{s_k\text{ for }k\in\mathcal{K}\}$ are linearly precoded as
\bea
\mathbf{x}_D=[\mathbf{x}_{D,1}^H\cdots\mathbf{x}_{D,N_R}^H]^H=\sum_{k\in\mathcal{K}}\mathbf{v}_{D,k}s_k,\label{x_D}
\eea
where $\mathbf{v}_{D,k}\in\mathbb{C}^{NN_R\times 1}$ is the $\textit{digital beamforming}$ vector across all the RRHs for the $k$th UE, and $\mathbf{x}_{D,i}\in\mathbb{C}^{N\times 1}$ represents the $i$th subvector of $\mathbf{x}_D\in\mathbb{C}^{NN_R\times 1}$ corresponding to the signal transmitted by the $i$th RRH.
Defining the shaping matrices $\mathbf{\Xi}_i=[\mathbf{0}_{N\times N(i-1)}^H~\mathbf{I}_N~\mathbf{0}_{N\times N(N_R-i)}^H]^H$, the $i$th subvector $\mathbf{x}_{D,i}$ can be expressed as $\mathbf{x}_{D,i}=\sum_{k\in\mathcal{K}}\mathbf{\Xi}_i^H\mathbf{v}_{D,k}s_k$.

Since the BBU communicates with the $i$th RRH via a fronthaul link of finite capacity, the signal $\mathbf{x}_{D,i}$ is quantized and compressed prior to being transferred to the RRH.
Following the approaches in \cite{OSi:16,SRL:18}, we model the impact of the compression by writing the quantized signal $\hat{\mathbf{x}}_{D,i}$ as
\bea
\hat{\mathbf{x}}_{D,i}=\mathbf{x}_{D,i}+\mathbf{q}_i,\label{hat_x_D}
\eea
where the quantization noise $\mathbf{q}_i\in\mathbb{C}^{N\times 1}\sim\mathcal{CN}(\mathbf{0},\mathbf{\Omega}_i)$ is independent of the signal $\mathbf{x}_{D,i}$.
As for the standard information-theoretic formulation, the covariance matrix $\mathbf{\Omega}_i$ describes the effect of the quantizer.
From \cite[Ch.~3]{AEGamal:12}, the quantized signal $\hat{\mathbf{x}}_{D,i}$ can be reliably recovered at the $i$th RRH, if the following condition is satisfied
\bea
&&\!\!\!\!\!\!\!\!\!\!\!\!g_i(\mathbf{V}_D,\mathbf{\Omega}_i)\triangleq I(\mathbf{x}_{D,i};\hat{\mathbf{x}}_{D,i})\label{g_i}\\
&&\!\!\!\!\!\!\!\!\!\!\!\!=\log_2\det\big(\sum_{k\in\mathcal{K}}\mathbf{\Xi}_i^H\mathbf{v}_{D,k}\mathbf{v}_{D,k}^H\mathbf{\Xi}_i\!+\!\mathbf{\Omega}_i\big)\!-\!\log_2\det(\mathbf{\Omega}_i)\!\leq\! C_i,\nonumber
\eea
where we define the set of the digital beamforming vectors as $\mathbf{V}_D\triangleq \{\mathbf{v}_{D,k}\text{ for }k\in\mathcal{K}\}$.

\subsection{RF Beamforming}
The quantized signal vector $\hat{\mathbf{x}}_{D,i}$ decompressed at the $i$th RRH is of dimension $N$, and is input to one of $N$ RF chains.
In order to fully utilize $M>N$ transmit antennas, the $i$th RRH applies $\textit{analog RF beamforming}$ to the signal $\hat{\mathbf{x}}_{D,i}$ via the beamforming matrix $\mathbf{V}_{R,i}\in\mathbb{C}^{M\times N}$.
The RF beamforming obtains $M$ signals for the antenna as a combination of $N$ output of the RF chains.
As a result, the transmitted signal $\mathbf{x}_i$ from $M$ transmit antennas is given as
\bea
\mathbf{x}_i=\mathbf{V}_{R,i}\hat{\mathbf{x}}_{D,i}=\sum_{k\in\mathcal{K}}\mathbf{V}_{R,i}\mathbf{\Xi}_i^H\mathbf{v}_{D,k}s_k+\mathbf{V}_{R,i}\mathbf{q}_i.\label{x_i}
\eea

As summarized in \cite{Eld:HF}, the RF beamforming can be implemented using analog phase shifters and switches.
Accordingly, each RF chain is connected to a specific set of transmit antennas through a phase shifter.
In this paper, we consider a fully connected phase shifter architecture, whereby each RF chain is connected to all transmit antennas via a separate phase shifter.
In the fully connected phase shifter architecture, the $(a,b)$th element of the RF beamforming matrix $\mathbf{V}_{R,i}$ is expressed as $\mathbf{V}_{R,i,a,b}=e^{j\theta_{i,a,b}}$ for $a\in\mathcal{M}$ and $b\in\mathcal{N}$, where $\theta_{i,a,b}$ indicates the phase shift between the signals $\hat{\mathbf{x}}_{D,i,b}$ and $\mathbf{x}_{i,a}$.
Therefore, when designing the RF beamforming matrix $\mathbf{V}_{R,i}$, one should satisfy constant modulus constraint $|\mathbf{V}_{R,i,a,b}|^2=1$ for $a\in\mathcal{M}$ and $b\in\mathcal{N}$ (see, e.g., \cite{FSo:16}).

\section{Design with Perfect CSI} \label{sec:design_perfect}
In this section, we discuss the problem of jointly designing the beamforming matrices for RF and digital beamforming, along with the fronthaul quantization noise covariance matrices.
As we will see, these problems are interdependent, since the impact of the quantization noise on the receivers' performance depends on the beamforming matrices.
Here, we first consider the case of perfect CSI, while the system with imperfect CSI will be addressed in Sec. \ref{sec:design_imperfect}.

To measure the achievable rate for each UE $k$, we rewrite the signal $y_k$ in (\ref{y_k}) under the transmission model (\ref{x_i}) as
\bea
y_k=\sum_{l\in\mathcal{K}}\mathbf{h}_k^H\bar{\mathbf{V}}_R\mathbf{v}_{D,l}s_l+\mathbf{h}_k^H\bar{\mathbf{V}}_R\mathbf{q}+z_k,\label{y_k2}
\eea
where $\bar{\mathbf{V}}_R\triangleq\text{diag}(\mathbf{V}_{R,i},\cdots,\mathbf{V}_{R,N_R})$ indicates the effective RF beamforming matrix across all RRHs and $\mathbf{q}\triangleq[\mathbf{q}_1^H\cdots\mathbf{q}_{N_R}^H]^H\in\mathbb{C}^{NN_R\times 1}$ stands for the vector of all the quantization noise signals, which is distributed as $\mathbf{q}\sim\mathcal{CN}(\mathbf{0},\bar{\mathbf{\Omega}})$ with $\bar{\mathbf{\Omega}}\triangleq\text{diag}(\mathbf{\Omega}_1,\cdots,\mathbf{\Omega}_{N_R})$.

Assuming that UE $k$ decodes the message $M_k$ by treating the interference signals as the additive noise, the achievable rate $R_k$ for the $k$th UE is given as
\bea
R_k&=&f_k(\mathbf{V}_R,\mathbf{V}_D,\mathbf{\Omega})=I(s_k;y_k)\label{R_k}\\
&=&\log_2\det(|\mathbf{h}_k^H\bar{\mathbf{V}}_R\mathbf{v}_{D,k}|^2+\zeta_k(\mathbf{V}_R,\mathbf{V}_D,\mathbf{\Omega}))\nonumber\\
&&-\log_2\det(\zeta_k(\mathbf{V}_R,\mathbf{V}_D,\mathbf{\Omega})),\nonumber
\eea
where $\mathbf{V}_R\triangleq\{\mathbf{V}_{R,i}\text{ for }i\in\mathcal{R}\}$, $\mathbf{\Omega}\triangleq\{\mathbf{\Omega}_i\text{ for }i\in\mathcal{R}\}$, and we denote the function $
\zeta_k(\mathbf{V}_R,\mathbf{V}_D,\mathbf{\Omega})\triangleq\sum_{l\in\mathcal{K}\backslash\{k\}}|\mathbf{h}_k^H\bar{\mathbf{V}}_R\mathbf{v}_{D,l}|^2+\mathbf{h}_k^H\bar{\mathbf{V}}_R\bar{\mathbf{\Omega}}\bar{\mathbf{V}}_R^H\mathbf{h}_k+\sigma_D^2$.

Considering the power consumption in the C-RAN, the total power consumption can be modeled as \cite{SRL:16}
\bea
P_T(\mathbf{V}_{R},\mathbf{V}_D,\mathbf{\Omega})&\triangleq&\sum_{i\in\mathcal{R}}p_i(\mathbf{V}_{R,i},\mathbf{V}_D,\mathbf{\Omega})\\
&&+N_U P_{N_U}+N N_R P_{RF},\nonumber
\eea
where the transmission power of the $i$th RRH is obtained as
\bea
&&\!\!\!\!\!\!\!\!p_i(\mathbf{V}_{R,i},\mathbf{V}_D,\mathbf{\Omega}_i)\triangleq\mathbb{E}\|\mathbf{x}_i\|^2\\
&&\!\!\!\!\!\!=\sum_{k\in\mathcal{K}}\text{tr}(\mathbf{V}_{R,i}\mathbf{\Xi}_i^H\mathbf{v}_{D,k}\mathbf{v}_{D,k}^H\mathbf{\Xi}_i\mathbf{V}_{R,i}^H)+\text{tr}(\mathbf{V}_{R,i}\mathbf{\Omega}_i\mathbf{V}_{R,i}^H),\nonumber
\eea
$P_{N_U}$ is the circuit power consumed by a UE, and $P_{RF}$ represents the circuit power consumption at each RRH, which is proportional to the number of RF chains.

\subsection{Weighted Sum-Rate Maximization} \label{sec:SR_perfect}
In this work, we tackle the problem of maximizing the WSR $\sum_{k\in\mathcal{K}}w_kR_k$ of the UEs while satisfying the per-RRH transmit power, fronthaul capacity and constant modulus constraints, where $w_k\geq0$ is a weight denoting the priority for the $k$th UE.
The problem is stated as
\begin{subequations}\label{P_1}
\begin{align}
\underset{\mathbf{V}_R,\mathbf{V}_D,\mathbf{\Omega}}{\text{maximize }}&\sum_{k\in\mathcal{K}}w_k f_k(\mathbf{V}_R,\mathbf{V}_D,\mathbf{\Omega})\label{P_1_ob}\\
\text{s.t.~}&g_i(\mathbf{V}_D,\mathbf{\Omega}_i)\leq C_i,~i\in\mathcal{R},\label{P_1_g_i}\\
&p_i(\mathbf{V}_{R,i},\mathbf{V}_D,\mathbf{\Omega}_i)\leq P_i,~i\in\mathcal{R},\label{P_1_p_i}\\
&|\mathbf{V}_{R,i,a,b}|^2=1,~a\in\mathcal{M},~b\in\mathcal{N},~i\in\mathcal{R}.\label{P_1_modul}
\end{align}
\end{subequations}

Problem (\ref{P_1}) is non-convex due to the objective function (\ref{P_1_ob}) and the constraints (\ref{P_1_g_i}), (\ref{P_1_p_i}) and (\ref{P_1_modul}).
In the next subsection, we present an iterative algorithm that computes an efficient solution of the problem.
To address problem (\ref{P_1}), we propose an iterative algorithm based on block coordinate descent (BCD), whereby the RF beamforming matrices $\mathbf{V}_R$ and the digital processing strategies $\{\mathbf{V}_D,\mathbf{\Omega}\}$ are alternately optimized.
We first describe the optimization of the digital part $\mathbf{V}_D$ and $\mathbf{\Omega}$ for a fixed RF beamforming $\mathbf{V}_R$, and then introduce the optimization of the latter.

\subsubsection{Optimization of Digital Beamforming and Fronthaul Compression} \label{sec:SR_digital_perfect}
For a given RF beamforming $\mathbf{V}_R=\mathbf{V}'_R$, problem (\ref{P_1}) with respect to the digital beamforming $\mathbf{V}_D$ and the fronthaul compression strategies $\mathbf{\Omega}$ can be written as
\begin{subequations}\label{P_D}
\begin{align}
\underset{\mathbf{V}_D,\mathbf{\Omega}}{\text{maximize }}&\sum_{k\in\mathcal{K}}w_k f_k(\mathbf{V}'_R,\mathbf{V}_D,\mathbf{\Omega})\label{P_D_ob}\\
\text{s.t.~}&g_i(\mathbf{V}_D,\mathbf{\Omega}_i)\leq C_i,~i\in\mathcal{R},\label{P_D_g_i}\\
&p_i(\mathbf{V}'_{R,i},\mathbf{V}_D,\mathbf{\Omega}_i)\leq P_i,~i\in\mathcal{R},\label{P_D_p_i}
\end{align}
\end{subequations}
where we eliminate the constant modulus constraint (\ref{P_1_modul}) which is independent of the digital variables $\mathbf{V}_D$ and $\mathbf{\Omega}$.
Problem (\ref{P_D}) is still non-convex due to the non-convex objective function (\ref{P_D_ob}) and constraint (\ref{P_D_g_i}).

To solve this problem, we extend the WMMSE-based algorithm in \cite{YZhou:16}.
To this end, we introduce two convex lower bounds on (\ref{P_D_ob}) and (\ref{P_D_g_i}) by applying a similar approach in \cite{YZhou:16}.
Denoting $x^{(\kappa)}$ as the quantity $x$ obtained at the $\kappa$th iteration of the BCD, a lower bound on the function $f_k(\mathbf{V}'_R,\mathbf{V}_D,\mathbf{\Omega})$ in (\ref{P_D_ob}) is written as
\bea
f_k(\mathbf{V}'_R,\mathbf{V}_D,\mathbf{\Omega})\geq\frac{1}{\ln2}\gamma_k(\mathbf{V}'_R,\mathbf{V}_D,\mathbf{\Omega},u_k^{(\kappa)},\tilde{w}_k^{(\kappa)}),\label{f_k}
\eea
where we define the function
\bea
&&\gamma_k(\mathbf{V}'_R,\mathbf{V}_D,\mathbf{\Omega},u_k^{(\kappa)},\tilde{w}_k^{(\kappa)})\label{gamma_k}\\
&&~~=\ln\tilde{w}_k^{(\kappa)}-\tilde{w}_k^{(\kappa)} e_k(\mathbf{V}'_R,\mathbf{V}_D,\mathbf{\Omega},u_k^{(\kappa)})+1,\nonumber
\eea
with arbitrary parameters $\tilde{w}_k^{(\kappa)}\geq0$ and $u_k^{(\kappa)}$, and the mean squared error (MSE) function is denoted as
\bea
&&e_k(\mathbf{V}'_R,\mathbf{V}_D,\mathbf{\Omega},u_k^{(\kappa)})\label{e_k}\\
&&~~=|1-(u_k^{(\kappa)})^*\mathbf{h}_k^H\bar{\mathbf{V}}'_R\mathbf{v}_{D,k}|^2+|u_k^{(\kappa)}|^2\zeta_k(\mathbf{V}'_R,\mathbf{V}_D,\mathbf{\Omega}).\nonumber
\eea
Note that the lower bound in (\ref{f_k}) is satisfied with equality when the variables $u_k^{(\kappa)}$ and $\tilde{w}_k^{(\kappa)}$ are equal to
\bea
&&u_k^{(\kappa)}=\tilde{u}_k(\mathbf{V}'_R,\mathbf{V}_D,\mathbf{\Omega})\\
&&~~~~~\triangleq\frac{\mathbf{h}_k^H\bar{\mathbf{V}}'_R\mathbf{v}_{D,k}}{|\mathbf{h}_k^H\bar{\mathbf{V}}'_R\mathbf{v}_{D,k}|^2+\zeta_k(\mathbf{V}'_R,\mathbf{V}_D,\mathbf{\Omega})},\label{u_k}\nonumber\\
&&\tilde{w}_k^{(\kappa)}=\frac{1}{e_k(\mathbf{V}'_R,\mathbf{V}_D,\mathbf{\Omega},u_k^{(\kappa)})}.\label{w_k}
\eea

Furthermore, an upper bound of the function $g_i(\mathbf{V}_D,\mathbf{\Omega}_i)$ in the constraint (\ref{P_D_g_i}) is given as
\bea
g_i(\mathbf{V}_D,\mathbf{\Omega})\leq\tilde{g}_i(\mathbf{V}_D,\mathbf{\Omega},\mathbf{\Sigma}_i^{(\kappa)}),\label{g_i_u}
\eea
for any arbitrary positive definite matrix $\mathbf{\Sigma}_i^{(\kappa)}$, where $\tilde{g}_i(\mathbf{V}_D,\mathbf{\Omega},\mathbf{\Sigma}_i^{(\kappa)})$ is represented as
\bea
&&\!\!\!\!\!\!\!\!\!\tilde{g}_i(\mathbf{V}_D,\mathbf{\Omega}_i,\mathbf{\Sigma}_i^{(\kappa)})=\log_2\det(\mathbf{\Sigma}_i^{(\kappa)})-\log_2\det(\mathbf{\Omega}_i)\label{g_i_t}\\
&&\!\!\!\!\!\!+\frac{\text{tr}\big((\mathbf{\Sigma}_i^{(\kappa)})^{-1}(\sum_{k\in\mathcal{K}}\mathbf{\Xi}_i^H\mathbf{v}_{D,k}\mathbf{v}_{D,k}^H\mathbf{\Xi}_i+\mathbf{\Omega}_i)\big)}{\ln2}-\frac{N}{\ln2}.\nonumber
\eea
Here, the matrix $\mathbf{\Sigma}_i^{(\kappa)}$ that achieves equality in (\ref{g_i_u}) is written as
\bea
\mathbf{\Sigma}_i^{(\kappa)}=\tilde{\mathbf{\Sigma}}_i(\mathbf{V}_D,\mathbf{\Omega})\triangleq\sum_{k\in\mathcal{K}}\mathbf{\Xi}_i^H\mathbf{v}_{D,k}\mathbf{v}_{D,k}^H\mathbf{\Xi}_i+\mathbf{\Omega}_i.\label{Sig}
\eea

Based on the inequalities (\ref{f_k}) and (\ref{g_i_u}), we formulate the problem
\begin{subequations}\label{P_D_}
\begin{align}
\!\!\!\!\!\underset{\mathbf{V}_D,\mathbf{\Omega},\mathbf{u}^{(\kappa)},\tilde{\mathbf{w}}^{(\kappa)},\mathbf{\Sigma}^{(\kappa)}}{\text{maximize }}&\sum_{k\in\mathcal{K}}\frac{w_k}{\ln2}\gamma_k(\mathbf{V}'_R,\mathbf{V}_D,\mathbf{\Omega},u_k^{(\kappa)},\tilde{w}_k^{(\kappa)})\label{P_D__ob}\\
\!\!\!\!\!\text{s.t.~}&\tilde{g}_i(\mathbf{V}_D,\mathbf{\Omega}_i,\mathbf{\Sigma}_i^{(\kappa)})\leq C_i,~i\in\mathcal{R},\label{P_D__g_i}\\
\!\!\!\!\!&p_i(\mathbf{V}'_{R,i},\mathbf{V}_D,\mathbf{\Omega}_i)\leq P_i,~i\in\mathcal{R},\label{P_D__p_i}
\end{align}
\end{subequations}
where $\mathbf{u}^{(\kappa)}\triangleq\{u_k^{(\kappa)}\text{ for }k\in\mathcal{K}\}$, $\tilde{\mathbf{w}}^{(\kappa)}\triangleq\{\tilde{w}_k^{(\kappa)}\text{ for }k\in\mathcal{K}\}$, and $\mathbf{\Sigma}^{(\kappa)}\triangleq\{\mathbf{\Sigma}_i^{(\kappa)}\text{ for }i\in\mathcal{R}\}$.
Although problem (\ref{P_D_}) is still non-convex, it is convex with respect to $\{\mathbf{V}_D,\mathbf{\Omega}\}$ when the variables $\{\mathbf{u}^{(\kappa)},\tilde{\mathbf{w}}^{(\kappa)},\mathbf{\Sigma}^{(\kappa)}\}$ are fixed and vice versa.
As proved in \cite{YZhou:16}, since each variable update yields a non-decreasing objective value in (\ref{P_D__ob}), solving problem (\ref{P_D_}) alternately over these two sets of variables would yield a solution that is guaranteed to converge to a stationary point.
This is detailed in Algorithm 1 below.

\renewcommand{\arraystretch}{1}
 \begin{center}
 \begin{tabular}{l}
 \hthickline
    $\bold{Algorithm~1}$: Algorithm for updating $\mathbf{V}_D$ and $\mathbf{\Omega}$\\
 \hthickline
    Set $\kappa=1$ and initialize $\mathbf{V}_D^{(\kappa)}$ and $\mathbf{\Omega}^{(\kappa)}$ satisfying the\\
    constraints (\ref{P_D_g_i})-(\ref{P_D_p_i}).\\
    $\bold{Repeat}$\\
    ~~~Update $u_k^{(\kappa)}=\tilde{u}_k(\mathbf{V}'_R,\mathbf{V}_D^{(\kappa)},\mathbf{\Omega}^{(\kappa)})$ for $k\in\mathcal{K}$.\\
    ~~~Update $\tilde{w}_k^{(\kappa)}=1/e_k(\mathbf{V}'_R,\mathbf{V}_D^{(\kappa)},\mathbf{\Omega}^{(\kappa)},u_k^{(\kappa)})$ for $k\in\mathcal{K}$.\\
    ~~~Update $\mathbf{\Sigma}_i^{(\kappa)}=\tilde{\mathbf{\Sigma}}_i(\mathbf{V}_D^{(\kappa)},\mathbf{\Omega}^{(\kappa)})$ for $i\in\mathcal{R}$.\\
    ~~~Update $\{\mathbf{V}_D^{(\kappa+1)},\mathbf{\Omega}^{(\kappa+1)}\}$ as a solution of problem (\ref{P_D_})\\
    ~~~for the given $\{\mathbf{u}^{(\kappa)},\tilde{\mathbf{w}}^{(\kappa)},\mathbf{\Sigma}^{(\kappa)}\}$.\\
    ~~~Set $\kappa\leftarrow \kappa+1$.\\
    $\bold{Until}$ convergence.\\
 \hthickline
 \end{tabular}
 \end{center}

\subsubsection{Optimization of RF Beamforming} \label{sec:SR_RF_perfect}
We now discuss the optimization of the RF beamformers $\mathbf{V}_R$ in problem (\ref{P_1}) for fixed digital variables $\mathbf{V}_D=\mathbf{V}'_D$ and $\mathbf{\Omega}=\mathbf{\Omega}'$.
The problem can be stated as
\begin{subequations}\label{P_A}
\begin{align}
\underset{\mathbf{V}_R}{\text{maximize }}&\sum_{k\in\mathcal{K}}w_k f_k(\mathbf{V}_R,\mathbf{V}'_D,\mathbf{\Omega}')\label{P_A_ob}\\
\text{s.t.~}&p_i(\mathbf{V}_{R,i},\mathbf{V}'_D,\mathbf{\Omega}'_i)\leq P_i,~i\in\mathcal{R},\label{P_A_p_i}\\
&|\mathbf{V}_{R,i,a,b}|^2=1,~a\in\mathcal{M},~b\in\mathcal{N},~i\in\mathcal{R}.\label{P_A_modul}
\end{align}
\end{subequations}

The presence of the constant modulus constraint (\ref{P_A_modul}) makes it difficult to solve problem (\ref{P_A}).
To address this issue, as in \cite[Sec.~III-A]{CSL:16}, we relax the condition (\ref{P_A_modul}) to the convex constraint $|\mathbf{V}_{R,i,a,b}|^2\leq 1$.
Then, the obtained problem can be solved by again applying the WMMSE method in \cite{YZhou:16}.
The procedure for solving problem (\ref{P_A}) is summarized in Algorithm 2, where the convex problem of the original problem (\ref{P_A}) is stated as
\begin{subequations}\label{P_A_1}
\begin{align}
       \underset{\mathbf{V}_R,\mathbf{u}^{(\kappa)},\tilde{\mathbf{w}}^{(\kappa)}}{\text{maximize }}&\sum_{k\in\mathcal{K}}\frac{w_k}{\ln 2}\gamma_k(\mathbf{V}_R,\mathbf{V}'_D,\mathbf{\Omega}',u_k^{(\kappa)},\tilde{w}_k^{(\kappa)})\\
       \text{s.t.~}&p_i(\mathbf{V}_{R,i},\mathbf{V}'_D,\mathbf{\Omega}'_i)\leq P_i,~i\in\mathcal{R},\label{P_A1_p_i}\\
       &|\mathbf{V}_{R,i,a,b}|^2\leq 1,~a\in\mathcal{M},~b\in\mathcal{N},~i\in\mathcal{R}.\label{P_A1_modul}
\end{align}
\end{subequations}

\begin{tabular}{l}
\hthickline
   $\bold{Algorithm~2}$: Algorithm for updating $\mathbf{V}_R$\\
\hthickline
   Set $\kappa=1$ and initialize $\mathbf{V}_R^{(\kappa)}$ satisfying the constraints\\
   (\ref{P_A_p_i})-(\ref{P_A_modul}).\\
   $\bold{Repeat}$\\
   ~~~Update $u_k^{(\kappa)}=\tilde{u}_k(\mathbf{V}_R^{(\kappa)},\mathbf{V}'_D,\mathbf{\Omega}')$ for $k\in\mathcal{K}$.\\
   ~~~Update $\tilde{w}_k^{(\kappa)}=1/e_k(\mathbf{V}_R^{(\kappa)},\mathbf{V}'_D,\mathbf{\Omega}')$ for $k\in\mathcal{K}$.\\
   ~~~Update $\mathbf{V}_R^{(\kappa)}$ as a solution of problem (\ref{P_A_1}) for the\\
   ~~~given $\{\mathbf{u}^{(\kappa)},\tilde{\mathbf{w}}^{(\kappa)}\}$.\\
   ~~~Set $\kappa\leftarrow \kappa+1$.\\
   $\bold{Until}$ convergence.\\
\hthickline
\end{tabular}

\vspace{10pt}

Since the RF beamforming matrices computed from Algorithm 2, denoted as $\tilde{\mathbf{V}}_{R}$, may not satisfy the constraint (\ref{P_A_modul}), we propose to obtain a feasible RF beamformer $\mathbf{V}_{R}$ by projecting $\tilde{\mathbf{V}}_{R}$ onto the feasible space \cite[Sec.~III-A]{CSL:16}.
Specifically, we find the RF beamformer $\mathbf{V}_{R,i}$ such that the distance $\|\mathbf{V}_{R,i}-\tilde{\mathbf{V}}_{R,i}\|^2_F$ is minimized.
As a result, the beamformer $\mathbf{V}_{R,i}$ is calculated as $\text{exp}(j\angle\tilde{\mathbf{V}}_{R,i,a,b})$ for $a\in\mathcal{M}$, $b\in\mathcal{N}$ and $i\in\mathcal{R}$ \cite[Eq. (14)]{CSL:16}.
In summary, for a joint design of the digital beamforming $\mathbf{V}_D$, the fronthaul compression $\mathbf{\Omega}$ and the RF beamforming strategies $\mathbf{V}_R$, we run Algorithm 1 and 2 alternately.
We note that while both Algorithm 1 and Algorithm 2 are individually convergent in the absence of modulus constraint for the RF beamforming,
due to the projection step in the update of RF beamforming, the overall alternating optimization algorithm is not guaranteed to converge.
This is also the case for the related algorithms in \cite{HG:16}.
Therefore, we will observe the convergence behavior of the proposed algorithm in Sec. \ref{sec:results}.

\subsection{Network Energy Efficiency Maximization}\label{sec:EE_perfect}
We now consider jointly designing RF and digital beamforming along with fronthaul compression with the aim of maximizing the overall network EE.
The network EE is defined as the ratio of the WSR to the corresponding power consumption.
Accordingly, the problem is formulated as
\begin{subequations}\label{P_e}
\begin{align}
\underset{\mathbf{V}_R,\mathbf{V}_D,\mathbf{\Omega}}{\text{maximize }}&
\frac{\sum_{k\in\mathcal{K}}w_k f_k(\mathbf{V}_R,\mathbf{V}_D,\mathbf{\Omega})}{P_T(\mathbf{V}_R,\mathbf{V}_D,\mathbf{\Omega})}\label{P_e_ob}\\
\text{s.t.~}&\text{(\ref{P_1_g_i}), (\ref{P_1_p_i}), (\ref{P_1_modul})}.\label{P_e_c}
\end{align}
\end{subequations}
Problem (\ref{P_e}) is also non-convex due to the objective function (\ref{P_e_ob}) and the constraints (\ref{P_e_c}).
In the following subsection, similar to Sec \ref{sec:SR_perfect}, we adopt alternating optimization to tackle problem (\ref{P_e}).

\subsubsection{Optimization of Digital Beamforming and Fronthaul Compression} \label{sec:EE_digital_perfect} \label{sec:EE_digital_perfect}
For a given RF beamforming $\mathbf{V}_R=\mathbf{V}'_R$, the digital beamforming $\mathbf{V}_D$ and the fronthaul compression strategies $\mathbf{\Omega}$ are optimized by solving the following problem
\begin{subequations}\label{P_e_D}
\begin{align}
\underset{\mathbf{V}_D,\mathbf{\Omega}}{\text{maximize }}&\frac{\sum_{k\in\mathcal{K}}w_k f_k(\mathbf{V}'_R,\mathbf{V}_D,\mathbf{\Omega})}{P_T(\mathbf{V}'_R,\mathbf{V}_D,\mathbf{\Omega})}\label{P_e_D_ob}\\
\text{s.t.~}&\text{(\ref{P_D_g_i}), (\ref{P_D_p_i})}.\label{P_e_D_c}
\end{align}
\end{subequations}

Since problem (\ref{P_e_D}) is non-convex, we also apply a similar approach proposed in Sec. \ref{sec:SR_digital_perfect}.
To make problem (\ref{P_e_D}) more tractable, we first introduce a new objective function as a natural logarithm of the objective function (\ref{P_e_D_ob})
\bea
\ln\big(\sum_{k\in\mathcal{K}}w_k f_k(\mathbf{V}'_R,\mathbf{V}_D,\mathbf{\Omega})\big)-\ln\big(P_T(\mathbf{V}'_R,\mathbf{V}_D,\mathbf{\Omega})\big).\label{P_e_D2_ob}
\eea
Then, we consider a convex lower bound of the function (\ref{P_e_D2_ob}) as
\bea
&&\!\!\!\!\!\!\!\!\!\!\!\!\ln(\sum_{k\in\mathcal{K}}w_k \label{L_1_EE} f_k(\mathbf{V}'_R,\mathbf{V}_D,\mathbf{\Omega}))-\ln(P_T(\mathbf{V}'_R,\mathbf{V}_D,\mathbf{\Omega}))\\
&&\geq \epsilon(\mathbf{V}'_R,\mathbf{V}_D,\mathbf{\Omega},\mathbf{u}^{(\kappa)},\tilde{\mathbf{w}}^{(\kappa)},\rho^{(\kappa)}),\nonumber
\eea
where we define the function
\bea
&&\!\!\!\!\!\!\!\!\!\!\!\!\!\!\!\!\!\!\epsilon(\mathbf{V}'_R,\mathbf{V}_D,\mathbf{\Omega},\mathbf{u}^{(\kappa)},\tilde{\mathbf{w}}^{(\kappa)},\rho^{(\kappa)})\\
&&\!\!\!\!\!\!\!\!=\ln\big(\sum_{k\in\mathcal{K}}\frac{w_k}{\ln2}\gamma_k(\mathbf{V}'_R,\mathbf{V}_D,\mathbf{\Omega},u_k^{(\kappa)},\tilde{w}_k^{(\kappa)})\big)-\ln(\rho^{(\kappa)})\nonumber\\
&&-\frac{P_T(\mathbf{V}'_R,\mathbf{V}_D,\mathbf{\Omega})}{\rho^{(\kappa)}}+1,\nonumber
\eea
with arbitrary parameters $u_k^{(\kappa)}$, $\tilde{w}_k^{(\kappa)}\geq0$ and $\rho^{(\kappa)}\geq0$.

One can show that for fixed $\{\mathbf{V}'_R,\mathbf{V}_D,\mathbf{\Omega}\}$, the lower bound in (\ref{L_1_EE}) holds with equality when the variables $\mathbf{u}^{(\kappa)}$, $\tilde{\mathbf{w}}^{(\kappa)}$ and $\rho^{(\kappa)}$ are given as
\bea
u_k^{(\kappa)}&=&\tilde{u}_k(\mathbf{V}'_R,\mathbf{V}_D,\mathbf{\Omega}),~k\in\mathcal{K},\\
\tilde{w}_k^{(\kappa)}&=&\frac{1}{e_k(\mathbf{V}'_R,\mathbf{V}_D,\mathbf{\Omega},u_k^{(\kappa)})},~k\in\mathcal{K},\\
\rho^{(\kappa)}&=&P_T(\mathbf{V}'_R,\mathbf{V}_D,\mathbf{\Omega}).
\eea

Based on the bounds (\ref{g_i_u}) and (\ref{L_1_EE}), the problem is formulated as
\begin{subequations}\label{P_e_R_con}
\begin{align}
\!\!\!\!\!\!\!\!\underset{\mathbf{V}_D,\mathbf{\Omega},\mathbf{u}^{(\kappa)},\tilde{\mathbf{w}}^{(\kappa)},\mathbf{\Sigma}^{(\kappa)},\rho^{(\kappa)}}{\text{maximize }}&\epsilon(\mathbf{V}'_R,\mathbf{V}_D,\mathbf{\Omega},\mathbf{u}^{(\kappa)},\tilde{\mathbf{w}}^{(\kappa)},\rho^{(\kappa)})\label{P_e2_ob}\\
\!\!\!\!\!\!\!\!\text{s.t.~}&\text{(\ref{P_D__g_i}), (\ref{P_D__p_i})}.
\end{align}
\end{subequations}
Similar to Algorithm 1, to obtain a solution $\{\mathbf{V}_D,\mathbf{\Omega}\}$, we alternately update the sets of variables $\{\mathbf{V}_D,\mathbf{\Omega}\}$ and $\{\mathbf{u}^{(\kappa)},\tilde{\mathbf{u}}^{(\kappa)},\mathbf{\Sigma}^{(\kappa)},\rho^{(\kappa)}\}$ until convergence.

\subsubsection{Optimization of RF Beamforming}
In this subsection, for fixed digital variables $\mathbf{V}_D=\mathbf{V}'_D$ and $\mathbf{\Omega}=\mathbf{\Omega}'$, we focus on optimizing the RF beamforming by solving the following non-convex problem
\begin{subequations}\label{P_e_R}
\begin{align}
\underset{\mathbf{V}_R}{\text{maximize }}&\frac{\sum_{k\in\mathcal{K}}w_k f_k(\mathbf{V}_R,\mathbf{V}'_D,\mathbf{\Omega}')}{P_T(\mathbf{V}_R,\mathbf{V}'_D,\mathbf{\Omega}')}\label{P_e_R_ob}\\
\text{s.t.~}&\text{(\ref{P_A_p_i}), (\ref{P_A_modul})}.\label{P_e_R_c}
\end{align}
\end{subequations}

To solve problem (\ref{P_e_R}), by using the bound (\ref{L_1_EE}) and relaxing the modulus constraint (\ref{P_A_modul}), we express the relaxed problem as
\begin{subequations}\label{P_e_R2}
\begin{align}
\underset{\mathbf{V}_R,\mathbf{u}^{(\kappa)},\tilde{\mathbf{w}}^{(\kappa)},\rho^{(\kappa)}}{\text{maximize }}&\epsilon(\mathbf{V}_R,\mathbf{V}'_D,\mathbf{\Omega}',\mathbf{u}^{(\kappa)},\tilde{\mathbf{w}}^{(\kappa)},\rho^{(\kappa)})\label{P_e_R2_ob}\\
\text{s.t.~}&\text{(\ref{P_A1_p_i}), (\ref{P_A1_modul})}.
\end{align}
\end{subequations}
Similar in Sec. \ref{sec:SR_RF_perfect}, the sets of variables $\mathbf{V}_R$ and $\{\mathbf{u}^{(\kappa)},\tilde{\mathbf{w}}^{(\kappa)},\rho^{(\kappa)}\}$ are alternately updated until convergence, and then the obtained RF beamforming matrices are projected onto the feasible space to satisfy the modulus constraint (\ref{P_A_modul}).
To sum up, the digital beamforming $\mathbf{V}_D$, the fronthaul compression $\mathbf{\Omega}$ and the RF beamforming $\mathbf{V}_R$ are jointly obtained by optimizing alternately $\{\mathbf{V}_D,\mathbf{\Omega}\}$ and $\mathbf{V}_R$.
The effectiveness of the proposed algorithm will be confirmed by numerical results in Sec. \ref{sec:results}.

\section{Design with Imperfect CSI} \label{sec:design_imperfect}
In the previous section, we have assumed that the instantaneous channel vectors $\mathbf{h}\triangleq\{\mathbf{h}_k\text{ for }k\in\mathcal{K}\}$ are perfectly known at the BBU.
In this section, we study a more practical case in which low-dimensional effective CSI $\{\bar{\mathbf{V}}_R^H\mathbf{h}_k\text{ for }k\in\mathcal{K}\}$ is acquired by the RRHs via uplink channel training in a TDD operation.
The key challenge is that the analog beamforming matrices affect the signal received on the uplink during the training phase.
Therefore, the design of the analog beamforming cannot rely on the knowledge of full CSI $\mathbf{h}$.
Instead, it is assumed that only the covariance matrices $\{\mathbf{R}_{k,i}\text{ for }k\in\mathcal{K},~i\in\mathcal{R}\}$ of the channel vectors are available at the BBU when designing analog precoding.
In practice, this long-term CSI can be estimated by means of time average if the fading channels are stationary for a sufficiently long time \cite{SPark:17,AAdhi:13}.

\subsection{Uplink Channel Training}\label{sec:training}
In the TDD operation, the downlink CSI is obtained based on the uplink training signals by leveraging reciprocity between downlink and uplink channels.
The channel matrix $\mathbf{H}_i=[\mathbf{h}_{1,i},\cdots,\mathbf{h}_{N_U,i}]\in\mathbb{C}^{M\times N_U}$ between all UEs and the $i$th RRH is estimated at the RRH and forwarded to the BBU.
Importantly, since channel estimation is performed based on the low-dimensional output of RF beamforming, the design of the RF beamforming $\mathbf{V}_R$ affects the channel estimation as well as the WSR performance.
For the rest of this subsection, we describe the relationships between $\mathbf{V}_R$ and the channel estimation error.

To elaborate, on the uplink, UE $k$ transmits the orthogonal training sequence $\psi_k\in\mathbb{C}^{L\times 1}$ of $L$ symbols with transmit power $p_k$, where the condition $L\geq N_U$ is required in order to ensure the orthogonality of the training sequences.
We have $\psi_k^H\psi_l=Lp_k\delta_{kl}$ for $k,l\in\mathcal{K}$, where $\delta_{ij}$ denotes the Kronecker delta function.
The signal matrix $\mathbf{Y}_i\in\mathbb{C}^{N\times L}$ received at the $i$th RRH during uplink training is given as
\bea
\mathbf{Y}_i=\mathbf{V}_{R,i}^H\mathbf{H}_i\mathbf{\Psi}^T+\mathbf{V}_{R,i}^H\mathbf{N}_i,
\eea
where $\mathbf{\Psi}=[\psi_1\cdots\psi_{N_U}]\in\mathbb{C}^{L\times N_U}$ represents the orthogonal training sequence matrix with $\mathbf{\Psi}^H\mathbf{\Psi}=\text{diag}(Lp_1,\cdots,Lp_{N_U})$ is the matrix of training signal powers, and $\mathbf{N}_i=[\mathbf{n}_{i,1}\cdots\mathbf{n}_{i,L}]\in\mathbb{C}^{M\times L}$ indicates the additive Gaussian noise matrix at the $i$th RRH with $\mathbf{n}_{i,l}\in\mathbb{C}^{M\times 1}\sim\mathcal{CN}(\mathbf{0},\sigma_U^2\mathbf{I}_M)$ for $l\in\{1,\cdots,L\}$.

To estimate the channel $\mathbf{H}_i$ from the received signal $\mathbf{Y}_i$, we define the received signal vector $\mathbf{y}_i\in\mathbb{C}^{NL\times 1}$ of the $i$th RRH as
\bea
\mathbf{y}_i&=&\text{vec}(\mathbf{Y}_i)\label{y_i}\\
&=&(\mathbf{\Psi}\otimes\mathbf{V}_{R,i}^H)\text{vec}(\mathbf{H}_i)+(\mathbf{I}_L\otimes \mathbf{V}_{R,i}^H)\text{vec}(\mathbf{N}_i),\nonumber
\eea
where $\text{vec}(\mathbf{X})$ denotes the vector obtained by stacking all columns of the matrix $\mathbf{X}$ on top of each other.
Note that the signal (\ref{y_i}) depends on the RF beamforming matrix $\mathbf{V}_{R,i}$.

Minimizing the MSE yields the estimated channel vector as
\bea
\hat{\mathbf{h}}_i=[\hat{\mathbf{h}}_{1,i}^H\cdots\hat{\mathbf{h}}_{N_U,i}^H]^H=\mathbf{W}_i\mathbf{y}_i,
\eea
where $\hat{\mathbf{h}}_{k,i}\in\mathbb{C}^{M\times 1}$ stands for the $k$th subvector of $\hat{\mathbf{h}}_i$ corresponding to the $k$th UE and $\mathbf{W}_i\triangleq\mathbf{R}_i(\mathbf{\Psi}^H\otimes\mathbf{V}_{R,i})((\mathbf{\Psi}\otimes\mathbf{V}_{R,i}^H)\mathbf{R}_i(\mathbf{\Psi}^H\otimes\mathbf{V}_{R,i})+(\mathbf{I}_L\otimes\sigma_U^2\mathbf{V}_{R,i}^H\mathbf{V}_{R,i}))^{-1}$ with $\mathbf{R}_i=\text{diag}(\mathbf{R}_{1,i},\cdots,\mathbf{R}_{N_U,i})$.
RRH $i$ sends the estimated channel vector $\hat{\mathbf{h}}_i$ to the BBU via the fronthaul link.
We assume that the coherence block is sufficiently large, so that the CSI overhead is amortized over many fronthaul channel uses.
As a result, the estimated channel vectors $\{\hat{\mathbf{h}}_i\text{ for }i\in\mathcal{R}\}$ are available at the BBU without additional distortion due to fronthaul transmission and does not violate the fronthaul capacity constraint \cite{JKang:14}.

\subsection{Weighted Sum-Rate Maximization}
We consider the problem of maximizing the average WSR of the UEs, while satisfying the per-RRH transmit power, fronthaul capacity and constant modulus constraints.
The problem is written as
\begin{subequations}\label{P_2}
\begin{align}
\!\!\!\!\!\underset{\mathbf{V}_R}{\text{maximize }}&\!\mathbb{E}_{\mathbf{h}}\!\big(\underset{\mathbf{V}_D(\mathbf{h}),\mathbf{\Omega}(\mathbf{h})}{\text{maximize }}\!\sum_{k\in\mathcal{K}}\!w_kf_k(\mathbf{V}_R,\mathbf{V}_D(\mathbf{h}),\mathbf{\Omega}(\mathbf{h}))\big)\label{P_2_ob}\\
\!\!\!\!\!\text{s.t.~}&g_i(\mathbf{V}_D(\mathbf{h}),\mathbf{\Omega}_i(\mathbf{h}))\leq C_i,~i\in\mathcal{R},~\forall\mathbf{h},\label{P_2_g_i}\\
\!\!\!\!\!&p_i(\mathbf{V}_{R,i},\mathbf{V}_D(\mathbf{h}),\mathbf{\Omega}_i(\mathbf{h}))\leq P_i,~i\in\mathcal{R},~\forall\mathbf{h},\label{P_2_p_i}\\
\!\!\!\!\!&|\mathbf{V}_{R,i,a,b}|^2=1,~a\in\mathcal{M},~b\in\mathcal{N},~i\in\mathcal{R}.\label{P_2_modul}
\end{align}
\end{subequations}
In problem (\ref{P_2}), we account for the fact that while the RF beamforming can only depend on long-term CSI, the digital beamforming and fronthaul compression can be a function of the instantaneous CSI.

\subsubsection{Design of RF Beamforming}\label{sec:design_SR_RF_imperfect}
As discussed before, the RF beamforming matrix affects both the downlink rate and the quality of the estimated CSI via uplink training.
In this subsection, we focus on the design of matrices $\mathbf{V}_R$ by assuming only the long-term CSI on the covariance matrices $\{\mathbf{R}_{k,i}\text{ for }k\in\mathcal{K},~i\in\mathcal{R}\}$.
Adopting the SAA method \cite{ALiu:16}, we generate $T$ instantaneous channel samples $\tilde{\mathbf{h}}\triangleq\{\tilde{\mathbf{h}}_t\text{ for }t\in\mathcal{T}\triangleq\{1,\cdots,T\}\}$ based on the second-order statistic $\{\mathbf{R}_{k,i}\text{ for }k\in\mathcal{R},i\in\mathcal{R}\}$ of the downlink channel vectors.
By approximating the objective function (\ref{P_2_ob}) with an empirical average as $\sum_{k\in\mathcal{K}}w_k \mathbb{E}_{\mathbf{h}}(f_k (\mathbf{V}_R,\mathbf{V}_D(\mathbf{h}),\mathbf{\Omega}(\mathbf{h})))\approx\sum_{k\in\mathcal{K}}\sum_{t\in\mathcal{T}}\frac{w_k}{T}f_k(\mathbf{V}_R,\mathbf{V}_D(\tilde{\mathbf{h}}_t),\mathbf{\Omega}(\tilde{\mathbf{h}}_t))$,
we formulate the problem as
\begin{subequations}\label{P_2_a}
\begin{align}
&\!\!\!\!\!\!\!\!\!\!\!\!\!\!\underset{\mathbf{V}_R,\mathbf{V}_D(\tilde{\mathbf{h}}),\mathbf{\Omega}(\tilde{\mathbf{h}})}{\text{maximize }}\sum_{k\in\mathcal{K}}\sum_{t\in\mathcal{T}}\frac{w_k}{T}f_k(\mathbf{V}_R,\mathbf{V}_D(\tilde{\mathbf{h}}_t),\mathbf{\Omega}(\tilde{\mathbf{h}}_t))\label{P_2_a_ob}\\
\text{s.t.~}&g_i(\mathbf{V}_D(\tilde{\mathbf{h}}_t),\mathbf{\Omega}_i(\tilde{\mathbf{h}}_t))\leq C_i,~i\in\mathcal{R},~t\in\mathcal{T},\label{P_2_a_g_i}\\
&p_i(\mathbf{V}_{R,i},\mathbf{V}_D(\tilde{\mathbf{h}}_t),\mathbf{\Omega}_i(\tilde{\mathbf{h}}_t))\leq P_i,~i\in\mathcal{R},~t\in\mathcal{T},\label{P_2_a_p_i}\\
&|\mathbf{V}_{R,i,a,b}|^2=1,~a\in\mathcal{M},~b\in\mathcal{N},~i\in\mathcal{R}.\label{P_2_a_modul}
\end{align}
\end{subequations}
Similar to problem (\ref{P_1}), we update the variables $\{\mathbf{V}_D(\tilde{\mathbf{h}}),\mathbf{\Omega}(\tilde{\mathbf{h}})\}$ from Algorithm 1 and the analog RF beamforming matrices $\mathbf{V}_R$ from Algorithm 2 alternately until convergence.

\subsubsection{Design of Digital Beamforming and Fronthaul Compression}
Based on the channel estimate obtained for the RF beamforming matrices $\mathbf{V}_R$ in Sec. \ref{sec:design_SR_RF_imperfect}, the BBU optimizes the digital strategies $\mathbf{V}_D$ and $\mathbf{\Omega}$.
We write the downlink received signal $y_k$ in (\ref{y_k2}) of the $k$th UE as
\bea
y_k=\sum_{l\in\mathcal{K}}(\hat{\mathbf{h}}_k^H\bar{\mathbf{V}}_{R}+\mathbf{e}_{k}^H)\mathbf{v}_{D,l}s_l+(\hat{\mathbf{h}}_k^H\bar{\mathbf{V}}_{R}+\mathbf{e}_k^H)\mathbf{q}+z_k,
\eea
where $\hat{\mathbf{h}}_k=[\hat{\mathbf{h}}_{k,1}~\cdots~\hat{\mathbf{h}}_{k,N_R}]\in\mathbb{C}^{MN_R\times 1}$ represents the estimated channel vector from all RRHs to the $k$th UE, and $\mathbf{e}_k\triangleq\mathbf{h}_k-\hat{\mathbf{h}}_k$ indicates the estimation error vector from all RRHs to the $k$th UE, which is distributed as $\mathbf{e}_k\sim\mathcal{CN}(\mathbf{0},\mathbf{E}_k)$ with $\mathbf{E}_k=\text{diag}(\mathbf{E}_{k,1},\cdots,\mathbf{E}_{k,N_R})$.
Here, the error covariance matrix $\mathbf{E}_{k,i}$ is given as
\bea
\mathbf{E}_{k,i}\!=\!\big(\mathbf{I}_M\!+\!\frac{Lp_k}{\sigma_U^2}\mathbf{R}_{k,i}\mathbf{V}_{R,i}(\mathbf{V}_{R,i}^H\mathbf{V}_{R,i})^{-1}\mathbf{V}_{R,i}^H\big)^{-1}\mathbf{R}_{k,i}.
\eea

Assuming that the estimation error and interference are treated as the additive noise \cite{BH:07}, the achievable rate for the $k$th UE is computed as
\bea
\!\!\!\!\!\!\!\!\bar{f}_k(\mathbf{V}_D,\mathbf{\Omega})&=&\log_2\det\big(|\hat{\mathbf{h}}_k^H\bar{\mathbf{V}}_R\mathbf{v}_{D,k}|^2+\bar{\zeta}_k(\mathbf{V}_D,\mathbf{\Omega})\big)\\
&&\!-\log_2\det(\bar{\zeta}_k(\mathbf{V}_D,\mathbf{\Omega})),\nonumber
\eea
where
\bea
&&\!\!\!\!\!\!\!\!\!\!\!\!\!\!\!\!\!\!\!\!\!\bar{\zeta}(\mathbf{V}_D,\mathbf{\Omega})\!\triangleq\!\!\!\!\sum_{l\in\mathcal{K}\backslash\{k\}}\!\!\!|\hat{\mathbf{h}}_k^H\bar{\mathbf{V}}_R\mathbf{v}_{D,l}|^2\!+\!\sum_{l\in\mathcal{K}}\!\!\mathbf{v}_{D,l}^H\bar{\mathbf{V}}_R^H\mathbf{E}_{k}\bar{\mathbf{V}}_R\mathbf{v}_{D,l}\\
&&~~~~+\hat{\mathbf{h}}_k^H\bar{\mathbf{V}}_R\bar{\mathbf{\Omega}}\bar{\mathbf{V}}_R^H\hat{\mathbf{h}}_k+\text{tr}(\bar{\mathbf{V}}_R^H\mathbf{E}_{k}\bar{\mathbf{V}}_R\bar{\mathbf{\Omega}})+\sigma_D^2.\nonumber
\eea

Then, the WSR maximization problem is formulated as
\begin{subequations}\label{P_22}
\begin{align}
\underset{\mathbf{V}_D,\mathbf{\Omega}}{\text{maximize }}&\sum_{k\in\mathcal{K}}w_k \bar{f}_k(\mathbf{V}_D,\mathbf{\Omega})\label{P_22_ob}\\
\text{s.t.~}&g_i(\mathbf{V}_D,\mathbf{\Omega}_i)\leq C_i,~i\in\mathcal{R},\label{P_22_g_i}\\
&p_i(\mathbf{V}_{R,i},\mathbf{V}_D,\mathbf{\Omega}_i)\leq P_i,~i\in\mathcal{R}.\label{P_22_p_i}
\end{align}
\end{subequations}
As in problem (\ref{P_D}), we adopt Algorithm 1 with minor modifications.

\subsection{Network Energy Efficiency Maximization}
We also address the problem of jointly optimizing RF and digital beamforming and fronthaul compression design with the goal of maximizing the average network EE.
The problem is written as
\begin{subequations}\label{Pe_2}
\begin{align}
      \underset{\mathbf{V}_R}{\text{maximize }}&\mathbb{E}_{\mathbf{h}}\bigg(\underset{\mathbf{V}_D(\mathbf{h}),\mathbf{\Omega}(\mathbf{h})}{\text{maximize }}\frac{\sum_{k\in\mathcal{K}}w_kf_k(\mathbf{V}_R,\mathbf{V}_D(\mathbf{h}),\mathbf{\Omega}(\mathbf{h}))}{P_T(\mathbf{V}_R,\mathbf{V}_D(\mathbf{h}),\mathbf{\Omega}(\mathbf{h}))}\bigg)\label{Pe_2_ob}\\
\text{s.t.~}&\text{(\ref{P_2_g_i}), (\ref{P_2_p_i}), (\ref{P_2_modul})}.
\end{align}
\end{subequations}

\subsubsection{Design of RF beamforming}\label{sec:design_EE_RF_imperfect}
Adopting the SAA method, the approximation problem of the original problem is written as
\begin{subequations}\label{Pe_22}
\begin{align}
\!\!\!\!\underset{\mathbf{V}_R,\mathbf{V}_D(\tilde{\mathbf{h}}),\mathbf{\Omega}(\tilde{\mathbf{h}})}{\text{maximize }}&\sum_{t\in\mathcal{T}}\frac{\sum_{k\in\mathcal{K}}w_kf_k(\mathbf{V}_R,\mathbf{V}_D(\tilde{\mathbf{h}}_t),\mathbf{\Omega}(\tilde{\mathbf{h}}_t))}{TP_T(\mathbf{V}_R,\mathbf{V}_D(\tilde{\mathbf{h}}_t),\mathbf{\Omega}(\tilde{\mathbf{h}}_t))}\label{Pe_22_ob}\\
\text{s.t.~}&\text{(\ref{P_2_a_g_i}), (\ref{P_2_a_p_i}), (\ref{P_2_a_modul})}.
\end{align}
\end{subequations}
Since the objective function (\ref{Pe_22_ob}) is a sum of ratios unlike the average WSR maximization problem, the proposed algorithm in Sec. \ref{sec:EE_perfect} cannot be directly applied to the average network EE maximization problem.
To tackle this issue, we introduce additional optimization variables $\boldsymbol{\alpha}\triangleq\{\alpha_t\text{ for }t\in\mathcal{T}\}$ which satisfy the non-convex constraint
\bea
\ln(\alpha_t)&\leq&\ln(\sum_{k\in\mathcal{K}}w_k f_k(\mathbf{V}_R,\mathbf{V}_D(\tilde{\mathbf{h}}_t),\mathbf{\Omega}(\tilde{\mathbf{h}}_t)))\label{EE_C1}\\
&&-\ln(P_T(\mathbf{V}_R,\mathbf{V}_D(\tilde{\mathbf{h}}_t),\mathbf{\Omega}(\tilde{\mathbf{h}}_t))),~t\in\mathcal{T}.\nonumber
\eea

Then, the problem can be equivalently recast as
\begin{subequations}\label{Pe_33}
\begin{align}
      \underset{\mathbf{V}_R,\mathbf{V}_D(\tilde{\mathbf{h}}),\mathbf{\Omega}(\tilde{\mathbf{h}}),\boldsymbol{\alpha}}{\text{maximize }}&\sum_{t\in\mathcal{T}}\frac{\alpha_t}{T}\label{Pe_33_ob}\\
\text{s.t.~}&\text{(\ref{P_2_a_g_i}), (\ref{P_2_a_p_i}), (\ref{P_2_a_modul}), (\ref{EE_C1})}.
\end{align}
\end{subequations}
However, since $\ln(\alpha_t)$ in (\ref{EE_C1}) is a concave function, it is still difficult to solve problem (\ref{Pe_33}).
To make the constraint (\ref{EE_C1}) more tractable, we also consider a convex upper bound on $\ln(\alpha_t)$ as
\bea
\ln(\alpha_t)\leq\ln(\beta_t^{(\kappa)})+\frac{\alpha_t}{\beta_t^{(\kappa)}}-1.\label{CC_1}
\eea

By using the convex bounds (\ref{g_i_u}), (\ref{L_1_EE}) and (\ref{CC_1}) and relaxing the modulus constraint (\ref{P_2_a_modul}), we formulate the problem
\begin{subequations}\label{Pe_44}
\begin{align}
&\!\!\!\!\!\!\!\!\!\!\!\underset{\mathcal{A}}{\text{maximize }}\sum_{t\in\mathcal{T}}\frac{\alpha_t}{T}\label{Pe_44_ob}\\
\text{s.t.~}&\tilde{g}_i(\mathbf{V}_D(\tilde{\mathbf{h}}_t),\mathbf{\Omega}_i(\tilde{\mathbf{h}}_t),\mathbf{\Sigma}_i^{(\kappa)}(\tilde{\mathbf{h}}_t))\!\leq\! C_i,\!~i\!\in\!\mathcal{R},\!~t\!\in\!\mathcal{T},\\
&p_i(\mathbf{V}_{R,i},\mathbf{V}_D(\tilde{\mathbf{h}}_t),\mathbf{\Omega}_i(\tilde{\mathbf{h}}_t))\leq P_i, i\in\mathcal{R},~t\in\mathcal{T},\\
&\epsilon(\mathbf{V}_R,\!\mathbf{V}_D(\tilde{\mathbf{h}}_t),\!\mathbf{\Omega}(\tilde{\mathbf{h}}_t),\!\mathbf{u}^{(\kappa)}\!(\tilde{\mathbf{h}}_t),\!\tilde{\mathbf{w}}^{(\kappa)}\!(\tilde{\mathbf{h}}_t),\!\rho^{(\kappa)}\!(\tilde{\mathbf{h}}_t))\\
&\geq\ln\beta_t^{(\kappa)}+\frac{\alpha_t}{\beta_t^{(\kappa)}}-1,~t\in\mathcal{T},\nonumber\\
&|\mathbf{V}_{R,i,a,b}|\leq1,~i\in\mathcal{R},~a\in\mathcal{M},~b\in\mathcal{N},
\end{align}
\end{subequations}
where $\mathcal{A}\triangleq\{\mathbf{V}_R,\mathbf{V}_D(\tilde{\mathbf{h}}),\mathbf{\Omega}(\tilde{\mathbf{h}}),\mathbf{u}^{(\kappa)}(\tilde{\mathbf{h}}),      \tilde{\mathbf{w}}^{(\kappa)}(\tilde{\mathbf{h}}),\mathbf{\Sigma}^{(\kappa)}(\tilde{\mathbf{h}}),$ $\rho^{(\kappa)}(\tilde{\mathbf{h}}),\boldsymbol{\alpha},\boldsymbol{\beta}\}$ with $\boldsymbol{\beta}\triangleq\{\beta_t\text{ for }t\in\mathcal{T}\}$.
As in Sec. \ref{sec:EE_perfect}, to obtain the RF beamforming matrices $\mathbf{V}_R$, we update the variables $\{\mathbf{V}_D(\tilde{\mathbf{h}}),\mathbf{\Omega}(\tilde{\mathbf{h}})\}$ and $\mathbf{V}_R$ alternately until convergence.

\subsubsection{Design of Digital beamforming and Fronthaul Compression}
After channel estimation based on the RF beamforming matrices $\mathbf{V}_R$ in Sec. \ref{sec:design_EE_RF_imperfect}, the digital beamforming $\mathbf{V}_D$ and the fronthaul compression strategies $\mathbf{\Omega}$ are computed by solving the following problem
\begin{subequations}\label{Pe_D1}
\begin{align}
\underset{\mathbf{V}_D,\mathbf{\Omega}}{\text{maximize }}&\frac{\sum_{k\in\mathcal{K}}w_k \bar{f}_k(\mathbf{V}_D,\mathbf{\Omega})}{P_T(\mathbf{V}_R,\mathbf{V}_D,\mathbf{\Omega})}\label{Pe_D1_ob}\\
\text{s.t.~}&\text{(\ref{P_22_g_i}), (\ref{P_22_p_i})}.
\end{align}
\end{subequations}
Problem (\ref{Pe_D1}) can be solved by the algorithm in Sec. \ref{sec:EE_digital_perfect} with minor modifications.
The performance of the proposed algorithm will be evaluated by numerical results in Sec. \ref{sec:results}.

\section{Numerical Results} \label{sec:results}
In this section, we present numerical results to validate the effectiveness of the proposed joint design of the RF and digital processing strategies.
Throughout the simulation results, we consider the case of $N_U=4$ UEs, $N_R=2$ RRHs, and $M=10$ RRH antennas, and evaluate the sum-rate of the UEs with $w_k=1$ for all $k\in\mathcal{K}$.
The length of pilot sequences is set to $L=N_U$, the uplink transmit power of all UEs is given as $p_k=1$ for all $k\in\mathcal{K}$, the circuit power per RF chain and the static power consumed at the UE are respectively fixed as $P_{RF}=1$ and $P_{N_U}=1$, and each RRH has the same fronthaul capacity $C$ and the same downlink transmit power $P$ for all $i\in\mathcal{R}$, i.e., $C_i=C$ and $P_i=P$.
In addition, we set the downlink noise variance to be $\sigma_D^2=1$ so that the downlink signal-to-noise ratio (SNR) is defined as $\text{SNR}=P$.

Following \cite{AAdhi:13,YJ:17}, we adopt the half wavelength-spaced uniform linear antenna array model of the RRH antennas such that
the channel covariance matrix $\mathbf{R}_{k,i}$ is given as
\bea
\mathbf{R}_{k,i,a,b}=\frac{1}{2\Delta_{k,i}}\int_{\theta_{k,i}-\Delta_{k,i}}^{\theta_{k,i}+\Delta_{k,i}}e^{-j\pi(a-b)\sin\phi}d\phi,
\eea
where the angle of arrival $\theta_{k,i}$ and the angular spread $\Delta_{k,i}$ have the distributions $\theta_{k,i}\sim \mathcal{U} (-\frac{\pi}{3},\frac{\pi}{3})$ and $\Delta_{k,i}\sim \mathcal{U} (\frac{\pi}{18},\frac{2\pi}{9})$, respectively.
Here, the notation $\mathcal{U}(a,b)$ represents a uniform distribution between $a$ and $b$.

\subsection{Perfect CSI Case}
For the perfect CSI case, we compare the performance of the proposed scheme in Sec. \ref{sec:design_perfect} with the following baseline schemes.
\begin{itemize}
\item \textit{Fully digital:} Fully digital beamforming is carried out across all RRH antennas, where beamforming is designed using Algorithm 1 with $M=N$.
\item \textit{Reduced-rank digital:} Fully digital beamforming is performed under rank constraint equal to $N$.
 This is done by running Algorithm 1 while omitting the projection step in the update of RF beamforming.
\item \textit{Random RF and optimized digital:} The phases of the RF beamforming matrices are randomly selected from an independent and identically distributed (i.i.d.) distribution $\mathcal{U}(0,2\pi)$.
\end{itemize}

\begin{figure}
\begin{center}
\includegraphics[width=2.95 in]{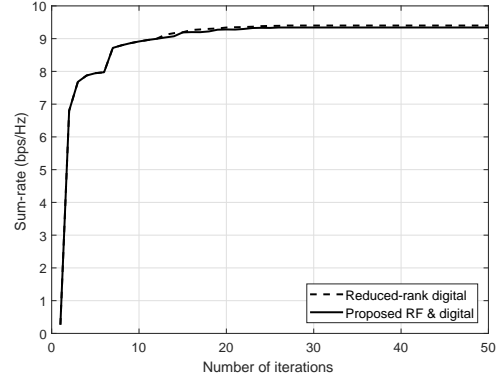}
\end{center}
\vspace{-5mm}
\caption{Convergence behavior for $N_U=4$, $N_R=2$, $N=2$, $M=10$, $C=5$ bps/Hz and $\text{SNR}=10$ dB.}
\label{figure:16}
\end{figure}

Fig. \ref{figure:16} illustrates the convergence behavior of the proposed algorithm for one channel realization with $C=5$ bps/Hz and $\text{SNR}=10$ dB.
The dashed line is obtained by the reduced-rank digital scheme, while the solid line is attained from the proposed algorithm.
The figure shows that in spite of the projection step, the proposed algorithm converges within a few tens of iterations.
In addition, the average per-iteration running time of the proposed algorithm and the fully digital scheme are 1.73 sec and 8.65 sec, respectively.
As can be seen, the number of RF chains is a important factor of the algorithm's running time.

\begin{figure}
\begin{center}
\includegraphics[width=2.95 in]{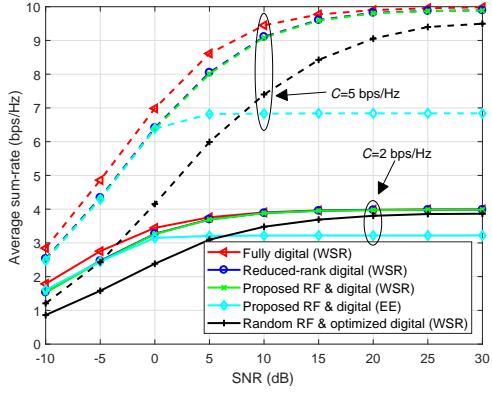}
\end{center}
\vspace{-5mm}
\caption{Average sum-rate performance with respect to SNR with $N_U=4$, $N_R=2$, $N=2$ and $M=10$.}
\label{figure:1}
\end{figure}

\begin{figure}
\begin{center}
\includegraphics[width=2.95 in]{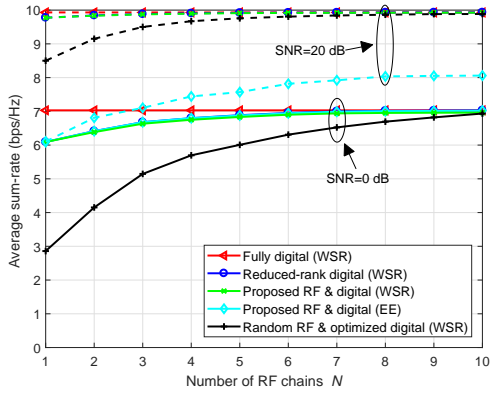}
\end{center}
\vspace{-5mm}
\caption{Average sum-rate performance with respect to $N$ with $N_U=4$, $N_R=2$, $M=10$ and $C=5$ bps/Hz.}
\label{figure:2}
\end{figure}

Fig. \ref{figure:1} shows the average sum-rate with respect to the downlink SNR for a C-RAN with $C\in\{2,5\}$ bps/Hz.
The proposed joint design of the RF and digital processing strategies always outperforms the random RF beamforming scheme, particularly at lower SNR, where the downlink channel sets the performance bottleneck of the system.
In a similar way, the optimization of RF beamforming has a more significant impact when the fronthaul capacity is larger.
As the SNR increases, the fronthaul capacity limitations become important, and thus the proposed joint design approaches the sum-rate of the fully digital scheme in spite of the limited number of RF chains.
Furthermore, by comparing the proposed WSR maximization and the reduced-rank digital scheme, we can check that the performance loss caused by the projection step in the update of RF beamforming is small.
In addition, it is seen that the sum-rate performance of the proposed EE maximization is consistent with the WSR approach at low SNR, but is saturated to a lower value at high SNR.
This is because in this regime the additional power needed to further increase the sum-rate is not necessary from the viewpoint of the EE.

In Fig. \ref{figure:2}, we plot the average sum-rate with respect to the number $N$ of RF chains for the downlink of a C-RAN with $C=5$ bps/Hz and $\text{SNR}\in\{0,20\}$ dB.
The sum-rate of the proposed joint design increases more rapidly with $N$ as compared to that of the random RF beamforming scheme.
Also, when $N$ is sufficiently large, both the proposed scheme and random RF beamforming achieve sum-rate performance very close to that of the fully digital beamforming scheme.
Similar to Fig. \ref{figure:1}, the impact of RF beamforming is more pronounced when the SNR is small for fixed fronthaul capacity.
It also confirms that the sum-rate performance gap between the proposed WSR maximization and EE maximization becomes larger as the SNR grows.

\begin{figure}
\begin{center}
\includegraphics[width=2.95 in]{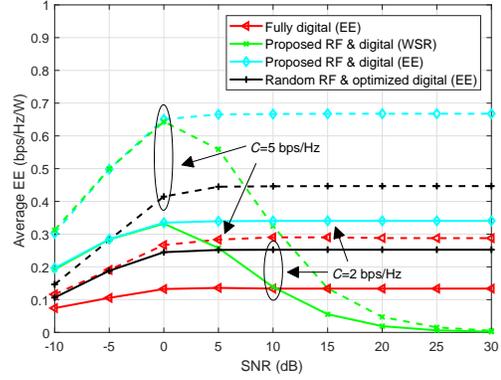}
\end{center}
\vspace{-5mm}
\caption{Average EE performance with respect to SNR with $N_U=4$, $N_R=2$, $N=2$ and $M=10$.}
\label{figure:3}
\end{figure}

Next, to investigate the EE performance of the proposed algorithm, Fig. \ref{figure:3} depicts the average EE with respect to the downlink SNR for $C\in\{2,5\}$ bps/Hz.
The figure illustrates the fact that as the SNR increases, WSR becomes extremely inefficient in terms of energy minimization.
This is in contrast to the schemes designed for EE maximization.
Furthermore, the fully digital architecture shows poor EE performance due to the energy consumed by $M$ RF chains at each RRH.

\begin{figure}
\begin{center}
\includegraphics[width=2.95 in]{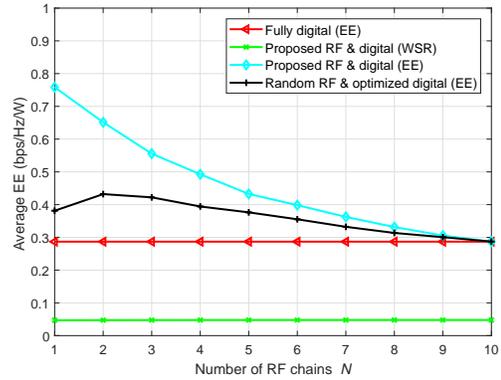}
\end{center}
\vspace{-5mm}
\caption{Average EE performance with respect to $N$ with $N_U=4$, $N_R=2$, $M=10$, $C=5$ bps/Hz and $\text{SNR}=20$ dB.}
\label{figure:4}
\end{figure}

This point is further explored in Fig. \ref{figure:4}, which shows the average EE with respect to the number $N$ of RF chains for $C=5$ bps/Hz and $\text{SNR}=20$ dB.
The main observation here is that increasing the number of RF chains may exhibit a negative impact on the EE, particularly when the RF beamforming is optimized.

\subsection{Imperfect CSI Case}

\begin{figure}
\begin{center}
\includegraphics[width=2.95 in]{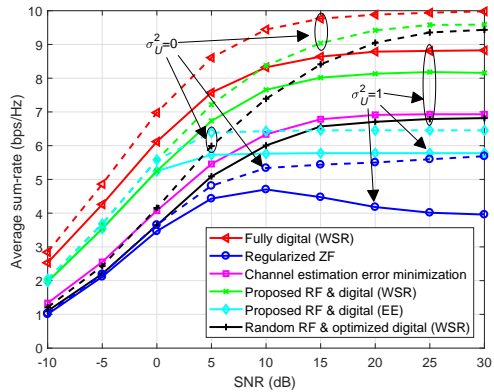}
\end{center}
\vspace{-5mm}
\caption{Average sum-rate performance with respect to SNR with $N_U=4$, $N_R=2$, $N=2$, $M=10$ and $C=5$ bps/Hz.}
\label{figure:5}
\end{figure}

\begin{figure}
\begin{center}
\includegraphics[width=2.95 in]{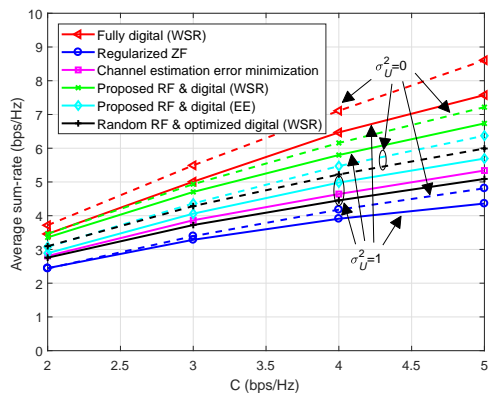}
\end{center}
\vspace{-5mm}
\caption{Average sum-rate performance with respect to $C$ with $N_U=4$, $N_R=2$, $N=2$, $M=10$ and $\text{SNR}=5$ dB.}
\label{figure:6}
\end{figure}

For the imperfect CSI case, we present the performance of two baseline schemes for comparison. The first is a regularized ZF method, whereby RF and digital beamforming is obtained from the algorithm in \cite{SPark:17} and fronthaul compression is determined using Algorithm 1 for given digital beamforming.
The second is a channel estimation error minimization scheme, where the RF beamforming matrices are computed by minimizing the MSE based on the algorithm in \cite{Eld:HF}, and the digital processing is calculated from Algorithm 1.

Fig. \ref{figure:5} plots the average sum-rate with respect to the SNR for a C-RAN with $C=5$ bps/Hz and $\sigma_U^2\in\{0,1\}$.
We consider the cases of a noiseless uplink channel $\sigma_U^2=0$ (dashed line) and a noisy uplink channel $\sigma_U^2=1$ (solid line).
In spite of the fact that the RF beamforming matrices are designed based on long-term CSI, we can check that the proposed WSR algorithm shows effective sum-rate performance.
In the presence of the channel estimation error $\sigma_U^2=1$, it is clear that the proposed digital beamforming is much more robust to the estimation errors in comparison to the regularized ZF scheme.
Also, although the impact of RF beamforming on the channel estimation error is considered, the channel estimation error minimization scheme is seen to yield suboptimal sum-rate performance.

In Fig. \ref{figure:6}, we illustrate the impact of the fronthaul capacity $C$ for a C-RAN with $\text{SNR}=5$ dB and $\sigma_U^2\in\{0,1\}$.
The proposed WSR scheme exhibits the average sum-rate which increases more rapidly with $C$ as compared to the random RF beamforming and regularized ZF schemes.
It is also observed that the proposed WSR scheme achieves the sum-rate performance close to the fully digital scheme at a small $C$.
In addition, the performance loss caused by the channel estimation error is relatively minor for the proposed WSR scheme as compared to the fully digital and random RF beamforming strategies, although the impact of RF beamforming on the channel estimation error is ignored.

\begin{figure}
\begin{center}
\includegraphics[width=2.95 in]{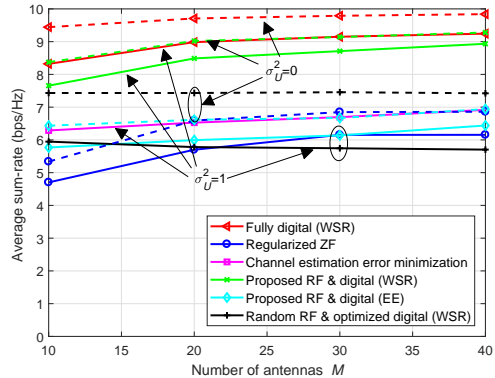}
\end{center}
\vspace{-5mm}
\caption{Average sum-rate performance with respect to $M$ with $N_U=4$, $N_R=2$, $N=2$, $C=5$ bps/Hz and $\text{SNR}=10$ dB.}
\label{figure:7}
\end{figure}

\begin{figure}
\begin{center}
\includegraphics[width=2.95 in]{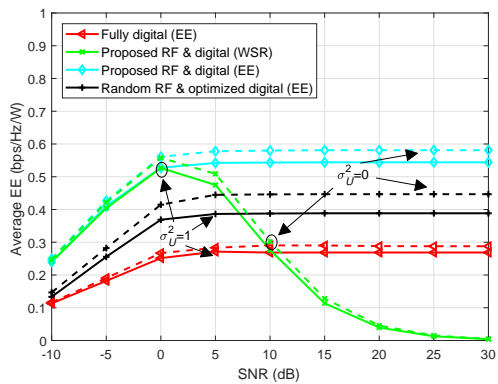}
\end{center}
\vspace{-5mm}
\caption{Average EE performance as a function of SNR with $N_U=4$, $N_R=2$, $N=2$, $M=10$ and $C=5$ bps/Hz.}
\label{figure:8}
\end{figure}

Fig. \ref{figure:7} depicts the average sum-rate in terms of the number $M$ of antennas for a C-RAN with $\text{SNR}=10$ dB, $C=5$ bps/Hz and $\sigma_U^2\in\{0,1\}$.
We can see that the average sum-rate performance gap between the fully digital scheme and the proposed WSR scheme becomes smaller as $M$ grows.
One interesting observation is that while the performance of the schemes with the optimized RF beamforming increases with $M$, the sum-rate of the random RF beamforming scheme decreases.
This implies that the optimization of RF beamforming is more significant when $M$ is larger.

Fig. \ref{figure:8} shows the average EE as a function of the SNR with $C=5$ bps/Hz and $\sigma_U^2\in\{0,1\}$.
Similar to Fig. \ref{figure:3}, we observe that when the SNR is sufficiently large, the proposed EE maximization scheme outperforms WSR.
Furthermore, the proposed joint design exhibits performance robust to channel estimation errors compared to randomized beamforming.

\section{Conclusions} \label{sec:conclusion}
We have studied the joint design for the downlink of a C-RAN with hybrid analog-digital antenna arrays.
Specifically, we have jointly optimized the digital beamforming, the fronthaul compression and the RF beamforming strategies with the goal of maximizing the WSR and the network EE, while satisfying the per-RRH power, fronthaul capacity and constant modulus constraints.
We have proposed an iterative algorithm that achieves an efficient solution under perfect CSI.
Furthermore, we have discussed the case of imperfect CSI based on the uplink channel training.
Numerical results have confirmed the effectiveness of the proposed algorithm.
Also, we have illustrated the impact of imperfect CSI on the downlink sum-rate and network EE performance and shown that the proposed scheme is robust to the estimation errors.
As some interesting directions for future researches, we mention the development of a globally optimal
algorithm and a design with low-resolution analog RF beamforming.
In addition, it will be interesting to consider joint multivariate compression also for this structure where the RRH focuses on analogue processing.
Furthermore, an uplink-downlink duality for this linear type pre-processing at the RRH remains as future work, where the idea is to extend the single user case \cite{LL:18} to multiple users.

\bibliographystyle{ieeetr}
\bibliography{AZREF}

\end{document}